\documentclass[a4paper]{jpconf}
\usepackage{graphicx}
\usepackage[dvips]{color}
\usepackage{inputenc}
\begin{document}
\title{Pseusogap in cuprates by Electronic Raman scattering}

\author{A. Sacuto, S. Benhabib, Y. Gallais, S. Blanc,  M. Cazayous, M.-A. M\'easson}

\address{Laboratoire Mat\'eriaux et Ph\'enom$\grave{e}$nes Quantiques (UMR 7162 CNRS),
Universit\'e Paris Diderot-Paris 7, Bat. Condorcet, 75205 Paris Cedex 13, France}

\author{J. S. Wen, Z. J. Xu and G. D. Gu}
\address {Matter Physics and Materials Science, Brookhaven National Laboratory (BNL), Upton, NY 11973, USA}

\ead{alain.sacuto@univ-paris-diderot.fr}

\begin{abstract}
We present Raman experiments on underdoped %($T_{c}= 75~K$, $p=0.11$) 
and overdoped $Bi_2Sr_2CaCu_2O_{8+\delta}$ (Bi-2212) single crystals. We reveal the pseudogap in the electronic Raman spectra in the $B_{1g}$ and $B_{2g}$ geometries. In these geometries we probe respectively, the antinodal (AN) and nodal (N) regions corresponding to the principal axes and the diagonal of the Brillouin zone. The pseudogap appears in underdoped regime and manifests itself in the $B_{1g}$ spectra by a strong depletion of the low energy electronic continuum as the temperature decreases. We define a temperature $T^*$ below which the depletion appears and the pseudogap energy, $\omega_{PG}$ the energy at which the depeletion closes.  

\par
The pseudogap is also present in the  $B_{2g}$ spectra but the depletion opens at higher energy than in the $B_{1g}$ spectra. We observe the creation of new electronic states inside the depletion as we enter the superconducting phase. This leads us to conclude (as proposed by S. Sakai et al. ~\cite{Sakai}) that the pseudogap has a different structure than the superconducting gap and competes with it.
We show that the nodal quasiparticle dynamic is very robust and almost insensitive to the pseudogap phase contrary to the antinodal quasiparticle dynamic. We finally reveal, in contrast to what it is usually admitted,an increase of the nodal quasiparticle spectral weight with underdoping. We interpret this result as the consequence of a possible Fermi surface disturbances in the doping range $p=0.1-0.2$. 

%The temperature dependence of the nodal static scatering rate is linear temperature dependent in the doping range of $p=0.11-0.22$. 

%of underdoped crystals but it manifests in a different manner than in $B_{2g}$ geometry . 
% (it increases linearly with temperature).
%In sharp contrast the low energy scattering rate (extracted form the $B_{2g}$ Raman response) exhibits a robust temperature dependence over a large doping level range ( from $p=0.11$ to $p=0.22$). This differs from the quadratic temperature dependence for the scatering rate detected from quantum oscillations measurements probably due to a too much lower doping level in our study. 

%We believe that our findings play a key role in the understanding of both the superconducting and the normal state 
 %up to nox is not taken into account by recent DMFT theory 
%and deserves to be considered in future theoretical investigations.

%We interpret this result as a possible consequence of the Fermi surface reconstruction in the doping range $p=0.1-0.2$. 
 
\end{abstract}

\section{Introduction}

The pseudogap phase has been first revealed by nuclear magnetic resonance (NMR) ~\cite{Warren,Alloul} and then it has been extensively studied by transport ~\cite{Ando,Loram,Tallon,Hussey,Taillefer}, angle resolved photoemission (ARPES)~\cite{Damascelli}, optics ~\cite{Homes,Timusk,Lobo} and scanning tunneling spectroscopy (STS)~\cite{Fischer}. In comparison, Raman scattering investigations of the pseudogap remain relatively scarce. One reason is that the pseudogap was first detected in the $B_{2g}$ geometry (nodal region) ~\cite{Nemetschek,Opel,Gallais} which introduced confusion with respect to ARPES data where the pseudogap sets in the antinodal region.
\par
In fact, Electronic Raman scattering investigations in the normal state have also revealed signatures of the pseudogap in $B_{1g}$ geometry (antinodal region) ~\cite{Blumberg,Guyard} however its effect has not been clearly quantified and well understood. For example the doping evolution of the pseudogap energy  $\omega_{PG}$ has not been studied in details and the doping level where the pseudogap disappears has not been determined. 
%versus doping has not been systematically defined from the Raman experiments in $B_{1g}$ and $B_{2g}$ geometries. % its energy and the pseudogap temperature T* no so well defined. 

% and (B2g) and few months later the Raman observation of a depletion of the electronic background in B1g channel 
Here, we present Raman experiments on underdoped and overdoped $Bi_2Sr_2CaCuO_{2+\delta}$ (Bi-2212) single crystals.
%propose focus on both  the B1g and and B2g geometries to track all the fingerprint of the pseudogap state in Bi2212 compound. 
In $B_{1g}$ geometry the pseudogap manifests itself by a strong depletion in the low energy Raman response function which develops as the temperature decreases. The pseudogap effect in $B_{2g}$ geometry is more subtle.  We detect a depletion in the Raman response function but at finite energy. At low energy the Raman response function exhibits a linear slope which increases with cooling as expected for a conventional metal~\cite{Opel}. Interestingly, we reveal that supplementary electronic states are created inside the energy range of the depletion as we enter in the superconducting state. We interpret our experimental findings as the unforseen $s-$wave anisotropic character of the pseudogap. This leads us to conclude that the pseudogap and the superconducting gap have different symmetries and most likely compete with each other~\cite{Sakai}.

%The low energy depletion detected at low temperature is filled up and disapears at high temperature.
%We have also properly identified  T* and the pseudogap energy. 

Finally we focus on the low energy Raman responses and extract the $\zeta=\Gamma/(Z^{*})^2$ ratio where $\Gamma$ is the static scattering rate and $Z^{*}$, the renormalized quasiparticle spectral weight~\cite{note}. 

%$\phi$ is the angle which defines a direction in the Brillouin zone with respect to its principal axes. 

In the antinodal region, $\zeta$ is temperature independent inside the pseudogap phase while it increases with temperature outside the pseudogap phase as expected for a metallic-like behavior for which the scattering rate has to increase with temperature.

In contrast in the nodal region $\zeta$ exhibits a metallic-like temperature dependence both inside and outside the pseudogap phase. 
%Interestingly its slope is larger for the overdoped sample than the underdoped one.
The equation of $\zeta$ depends both on $Z^{*}$ and $\Gamma$ which are difficult to disentangle in a first approach. However, a simple modeling of the Raman response function in Drude like conductivity allows us to follow separately the doping and temperature dependences of these two physical quantities in both the nodal and antinodal regions.  We show that $Z_{N}^{*}$ exhibits an unexpected increase with underdoping  while $Z_{AN}^{*}$ decreases. Simultaneously the static scattering rate $\Gamma_{AN}$ strongly increases while the  $\Gamma_{N}$ exhibits only small changes. We interpret this result as a consequence of a Fermi surface disturbances  between $0.11$ and $0.22$ in which well defined quasiparticles are preserved and even enhanced in the nodal region while they are destroyed in the antinodal region as the doping level is reduced. We find that $\Gamma_{N}$ exhibits a quasi linear temperature dependence in and out of the pseudogap phase. 

\section{Experimental Results}

The Bi-2212 single crystals have been grown by floating zone method. Doping is achieved by changing the oxygen content only. The detailed procedures of the crystal growth are described elsewhere ~\cite{Gu,Mihaly}. Raman experiments have been carried out using a triple grating spectrometer (JY-T64000) equipped with a nitrogen cooled CCD detector. We used the 532 nm excitation line from a diode pumped solid state laser (DPSS) and the 514.52 nm line from a $Ar^{+}$,$Kr^{+}$ laser. Temperature dependent measurements have been carried out in a ARS closed cycle cryostat.  All measurements have been corrected for the Bose factor and the instrumental spectral response. The $B_{1g}$ and $B_{2g}$ geometries have been obtained using crossed light polarizations at 45$^o$ from the $Cu-O$ bond directions and along them respectively. 
%In these geometries we probe respectively, the antinodal (AN) and nodal (N) regions corresponding to the principal axes and the diagonal of the Brillouin zone.
\par
Special care has been devoted to make reliable quantitative comparisons between the Raman intensities of distinct crystals with different doping levels measured in the same geometry, and between measurements in distinct geometries for crystals with the same doping level.  We have performed all the measurements during the same run and the crystals with various doping levels have been mounted on the same sample holder in order to keep the same optical configuration. The optical constant, deduced from spectroscopic ellipsometry measurements on Bi-2212 single crystals, have also been taken into account in order to have comparable Raman intensities of crystals with distinct doping levels.

%Finally, the Raman cross-section at each doping level was obtained by correcting the Raman response function for the optical constants, using the following expression for the correction factor :
%$\frac{[\alpha(\omega_{s})+\alpha(\omega_{i})].n(\omega_{s})^2}{T(\omega_{s}).T(\omega_{i})}$ where $\alpha$ referred to the absorption, $T$ to the transmission at the air-sample interface and $n$ to the complex refractive index whose components have been determined from spectroscopic ellipsometry measurements.  

\begin{figure}[ht!]
\begin{center}
\includegraphics[width=14cm]{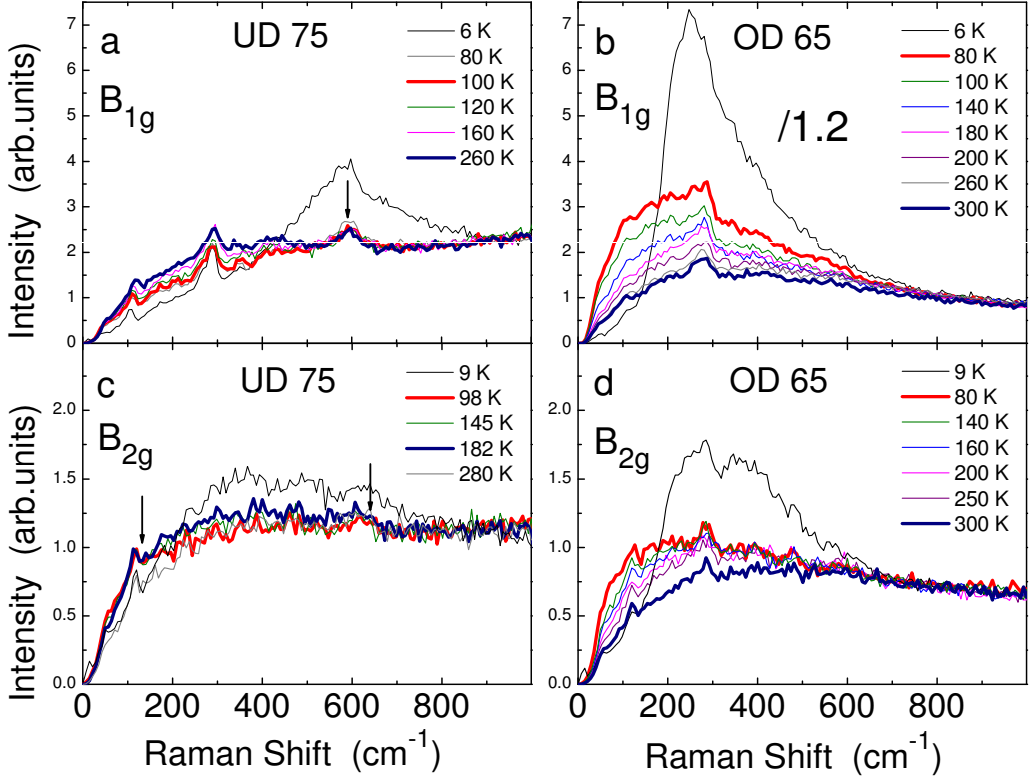}
\end{center}\vspace{-7mm}
\caption{$B_{1g}$ and $B_{2g}$ Raman response functions of underdoped and overdoped Bi-2212 single crystals. Spectra have been measured with the 532 nm laser line. The UD 75 and OD 65 single crystals have been measured after two successive exfoliations. In pannel b, the Raman intensities have been divided by 1.2} 
\label{fig1}
\end{figure}

In Figure 1 are displayed the temperature dependence of the $B_{1g}$ and $B_{2g}$ Raman response functions for two distinct Bi-2212 single crystals. The first one is an underdoped (UD) crystal $p=0.11$ with a $T_c=75~K$ and the second one is overdoped (OD) $p=0.22$ with $T_c=65~K$. The Raman response functions in the superconducting state (black thin curves) exhibit a pair breaking peak. This peak is much more pronounced in the $B_{1g}$  than in $B_{2g}$  geometry (for the same doping level) as expected for a d-wave superconducting gap ~\cite{Devereaux}. We can also notice that the $B_{1g}$ superconducting peak increases in intensity while its energy decreases with doping.  This has already been pointed out in our previous works ~\cite{Blanc1,Sacuto} and we have assigned the integrated area under the $B_{1g}$ pair breaking peak to the density of Cooper pairs in the antinodal region.
In the normal state (Fig.1-a) the $B_{1g}$  Raman response function of the UD 75 crystal exhibits a strong depletion of the low frequency  electronic continnum as the temperature decreases. The thick curves underline the low energy depletion of the electronic background in the Raman spectra. It extends from the lowest energy to approximatively $600~cm^{-1}$ (see arrow). We define the end of this depletion in the spectra as the pseudogap energy $\omega_{PG}$. 
Such a low energy depletion has been qualitatively reproduced by recent cluster dynamical mean field methods~\cite{Sakai,Lin}. 
\par

\begin{figure}[ht!]
\begin{center}
\includegraphics[width=10cm]{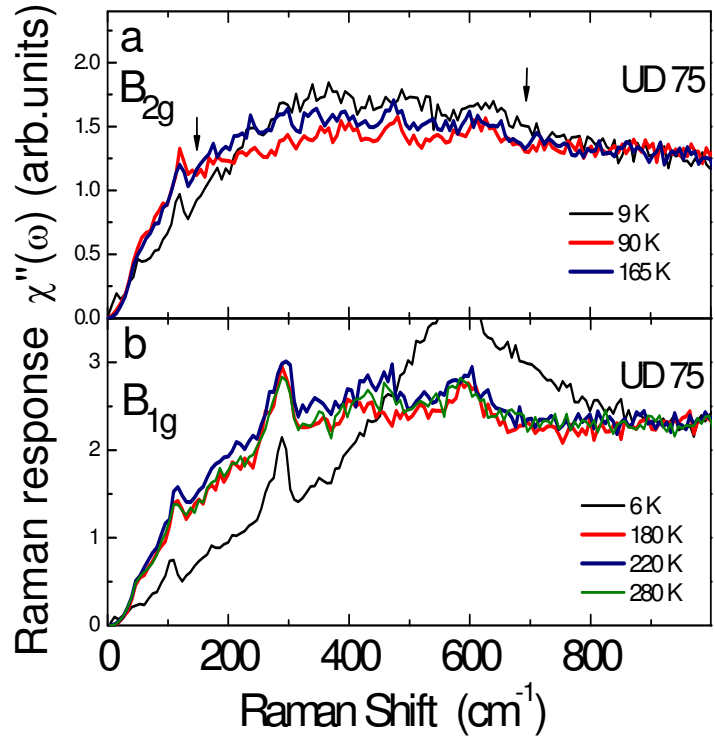}
\end{center}\vspace{-7mm}
\caption{Selected (a) $B_{2g}$ and (b) $B_{1g}$ Raman spectra of UD 75 compound.} 
\label{fig2}
\end{figure}

The non trivial fact is that a depletion is also detected in the $B_{2g}$ geometry where it opens at finite energy, above $150~cm^{-1}$, and closes around $600~cm^{-1}$ (see arrows in Fig.1-c and 2-a). This has also been reported in previous works \cite{Nemetschek,Opel,Gallais}. Below $150~cm^{-1}$, the $B_{2g}$ Raman response is not altered by the depletion and the slope increases as the temperature decreases. This is expected for a metal since the low energy slope of the Raman response is proportional to the quasiparticle lifetime~\cite{Devereaux} . 
%but a depletion sets in  at finite energy above (150cm-1) (quoted by an arrow) and close around 650cm-1, see fig 2a,  
 %the depletion developps with cooling. 
In sharp contrast, no trace of depletion is detected in the Raman spectra of the OD 65 crystal in both $B_{1g}$ and $B_{2g}$ geometries (see figs 1-b,d) and the low energy slopes of the Raman response functions increase with cooling as expected for a metal \cite{Devereaux,Venturini}.

\begin{figure}[ht!]
\begin{center}
\includegraphics[width=10cm]{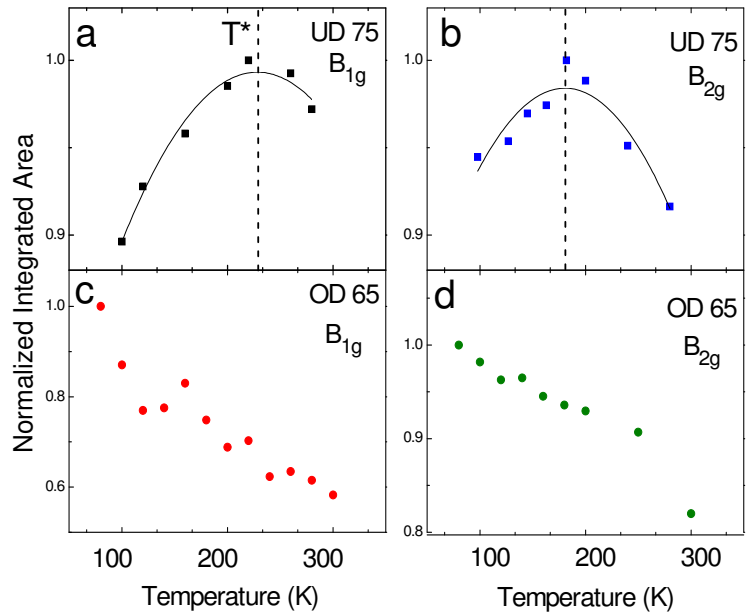}
\end{center}\vspace{-7mm}
\caption{Temperature dependence of the normalized integrated area of the Raman response functions shown in fig.1. Normalization has been achieved by dividing the areas by their maximum value. The vertical dashed line indicates $T^{*}$ and the thin line corresponds to a polynomial fit.} 
\label{fig3}
\end{figure}

\par

We can estimate the pseudogap temperature $T^*$ by studying the temperature dependence of the normalized integrated area of the $B_{1g}$ Raman response (up to $800~cm^{-1}$). The integrated area are plotted in Fig. 3. In fig. 3-a, (ud 75),  the positive slope corresponds to a filling of the depletion as the temperature increases.  The electronic background is then progressively restored and we can define the temperature $T^*$ for which the sign of the slope changes. It is approximatively $230~K$ (see dashed line).  
%and 175K in B2g channel for the UD 75K crystal. 
Beyond, $T^*$ the integrated electronic continuum starts to decrease. The slope of the integrated area as a function of temperature is negative. This is a consequence of the decrease of the low energy slope of the Raman response with temperature. This is can be seen in fig.2-b where the low energy slope of the $B_{1g}$ Raman response decreases between $220$ and $280~K$. This corresponds to a metallic behavior ~\cite{Opel,Devereaux,Venturini}. 
\par
The $B_{2g}$ integrated area of UD 75 exhibits a similar temperature behaviour as shown in fig. 3-b.  However, the extracted $T^*$ value is smaller than the $B_{1g}$ one. A possible reason for this behaviour is that the $B_{2g}$ depletion opens at finite energy while the low electronic continuum still exhibit a metallic behaviour. The low energy continuum thus decreases as temperature raises playing against the filling of the depletion. This effect may reduce the extracted  $T^*$ in the nodal region with respect to the antinodal one.(see ref. ~\cite{Sakai}).
% range than the B1g one and opens at finite energy and oneaIt  and a metallic behaviour is seen at lower energy. This reduces the variation amplitude of integrated area with respect to the  $B_{1g}$ one. We can also notice that $T^*_N$ is smaller than  $T^*_AN$.  
%with respect to the one in the $B_{1g}$ geometry. 
In the following we will assume that the value of $T^*$ directly related to the pseudogap temperature is the $B_{1g}$ one.  

%that that the low energy slope of the electronic background as a consequence the  the B2g integrated area as a fucntion of temperature is smaller than in B1g geometry and T* is weaker. 
\par
In the overdoped case (see fig.3-c,d) the temperature dependence of the integrated area in $B_{1g}$ and $B_{2g}$ geometries exhibit negative slopes down to the lowest temperatures. This corresponds to a continuous intensity decrease of the electronic background with temperature as expected for a metal\cite{Devereaux}. Our experimental findings indicate that the pseudogap phase only sets in below $p=0.22$.

\begin{figure}[ht!]
\begin{center}
\includegraphics[width=12cm]{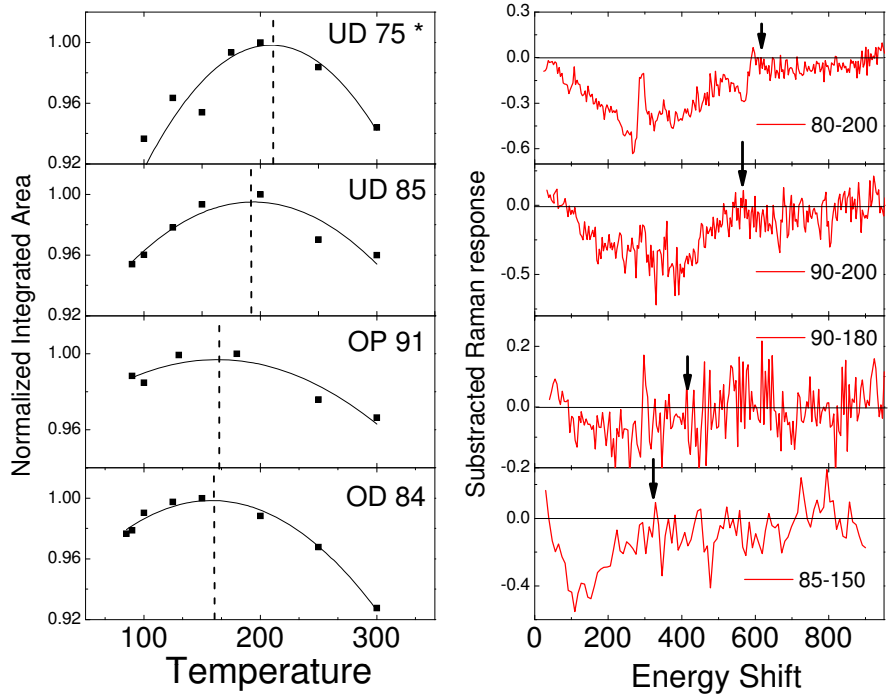}
\end{center}\vspace{-7mm}
\caption{left panel : Normalized integrated area for the $B_{1g}$ Raman spectra of Bi-2212 crystals. Normalization has been made from the maximum. The vertical dashed line indicates $T^{*}$ and the thin line corresponds to a polynomial fit; right panel: Subtracted Raman responses measured close to $T^*$ and just above $T_{c}$. Measurements have been performed with the 514.52 nm laser line. The UD 75 single crystal has been measured without exfoliation.} 
\label{fig4}
\end{figure}

In order to map $T^*$ versus doping, we have performed Raman measurements on Bi-2212 crystals with various doping levels. This is shown in the left panel of fig.4  where the integrated area of the $B_{1g}$ Raman response function up to $800~cm^{-1}$ is plotted for various dopin levels. The dash lines correspond to the temperature from which the slope of the integrated area curve changes of sign. 
%This gives us a more precise estimate for $T^*$ value. 
We observe that $T^*$ increases with underdoping. We have also defined the pseudogap energy by subtracting the Raman response measured close to $T^*$  from the one just above $T_c$. The subtracted spectra are shown in the right panel of fig.4. The pseudogap energy $\omega_{PG}$ corresponds to the energy for where the depletion closes. We see that $\omega_{PG}$ also increases with underdoping.

\begin{figure}[ht!]
\begin{center}
\includegraphics[width=10cm]{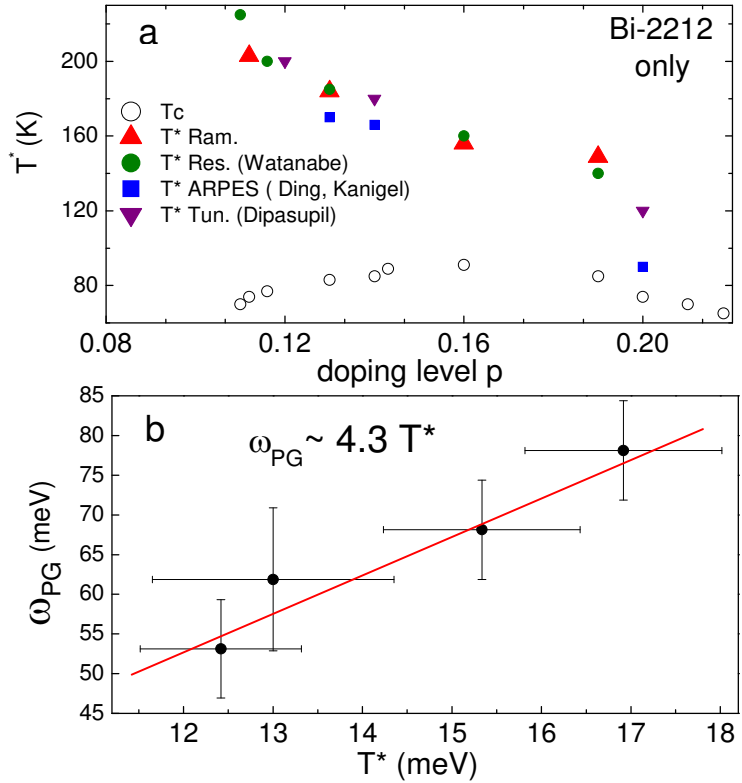}
\end{center}\vspace{-7mm}
\caption{(a) Doping evolution of the pseudogap temperature $T^*$ obtained from various techniques on Bi-2212 compound.(b)$\omega_{PG}$ versus $T^*$} 
\label{fig5}
\end{figure}

\par

We have also plotted the pseudogap energy obtained from distinct probes on the same Bi-2212 system: resistivity~\cite{Watanabe}, ARPES ~\cite{Ding,Kanigel} tunneling~\cite{Dipasupil} and our Raman data in $B_{1g}$ geometry. We can see in fig. 5a that all the measurements are consistent each other. We observe that  $T^*$ exhibits the same doping evolution and the same order of magnitude. Typically $T^*$ increases from $100~K$ to $230~K$ between $p=0.1$ and $0.2$. This makes us confident about our procedure to define $T^*$. We have also plotted in fig.5-b $\omega_{PG}$ versus $T^*$ and we find $\omega_{PG}\approx 4 T^*$ as it was observed from tunneling \cite{Kugler}. 

\section{Two Distinct Structures for the Pseudo gap and the Supercoducting gap} 
\par
We now discuss the implications of our findings on the pseudogap structure and its symmetry with respect to the superconducting gap. 

%The Temprature dependence of the B1g Raman response function have qualitatively repoduced by several group using 'cluster' dynamical mean field theory ( cite Lin et al PRB 82, 0455104, 2010 and S.Sakai). 
The starting point is the unusual depletion of the $B_{2g}$ Raman response which sets in an intermediate energy range. This has recently suggested a new interpretation for the pseudogap structure ~\cite{Sakai} deduced from cellular dynamical mean field theory (CDMFT) \cite{Kotliar}. The pseudogap would be distinct from the $d-$wave symmetry ~\cite{Sakai}. Indeed the opening of a $d-$wave pseudogap around the antinodes (as sketches in fig.6-a) should generate a continuous loss of electron-hole pair excitations in a large energy range from very low to high energy even in nodal region. Inside the k-window fixed by the angular extension of the $B_{2g}$ Raman vertex, very low lying energy excitations are allowed but there is no simple way to get a depletion for intermediate energy range only.  In contrast an s-wave anisotropic pseudogap which sets in away from the Fermi surface in the nodal region, like the one depicted in fig. 6-b, leads naturally to a depletion at finite energy. This is because low energy excitations are allowed in the nodal region while the intermediate energy ones are forbidden until the pseudogap energy is exceeded (see fig.4-b). Our view is consistent with recent ARPES data where particle-hole asymmetry has been found in the nodal region. It was obtained on states thermally occupied above the Fermi level~\cite{Johnson}

\begin{figure}[ht!]
\begin{center}
\includegraphics[width=8cm]{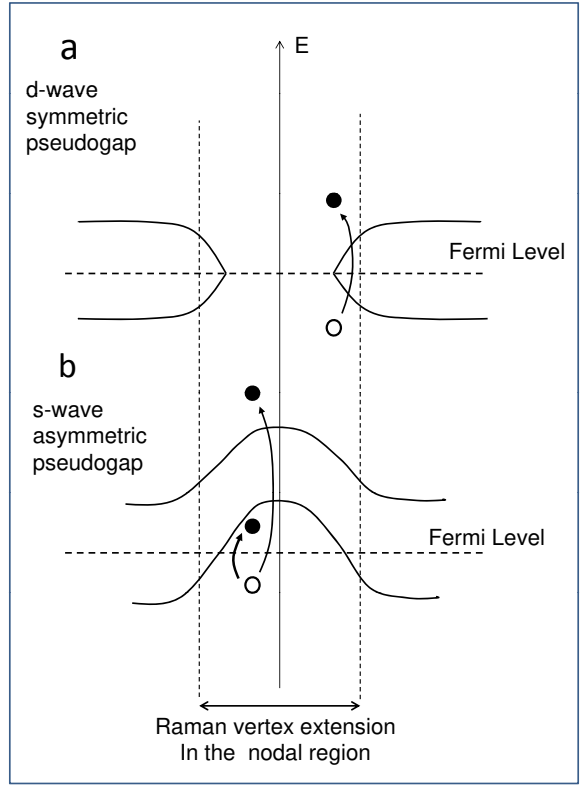}
\end{center}\vspace{-7mm}
\caption{(a) d-wave symmetric and (b) s-wave asymmetric shapes of the pseudogap in the nodal region. electron-hole pair excitations are sketched for the both scenarios. The band structure below the Fermi level looks similarly between $d$ and $s$-wave pseudogaps but are different above the Fermi level. } 
\label{fig6}
\end{figure}

The second crucial point that our data reveal is the presence of additional electronic states generated inside the pseudogap when we enter the superconducting state.  In the $B_{2g}$ geometry,(see fig. 2-a)  the superconducting peak emerges inside the energy range $150-600~cm^{-1}$ where the depletion sets in. This is also observed in the  $B_{1g}$ geometry (see fig. 2-b). The low energy edge of the superconducting peak is located near $400~cm^{-1}$ inside the energy range of the depletion. Creation of new states within the pseudogap is in agreement with recent advances in dynamical mean field theory which propose a competition between the superconducting gap and the pseudogap\cite{Sakai,Millis2,Sordi}.
\par
These two experimental facts make very improbable a performed pairs scenario for the pseudogap phase.  The first reason is that we should have the same symmetry for both the pseudogap and the superconducting gap. This is not the case here. The second reason is that we should expect an enhancement of the  depletion in the superconducting state if this last one results from preformed pairs. This is clearly not the case since the pair breaking peak emerges inside the depletion. 
Consequently, our experimental findings show that the pseudogap and superconducting gap are distinct in origin. This answers a long standing question introduced by Norman in 2005\cite{Norman}: ''The pseudogap: friend or foe of high Tc?''. It is a foe.
Our findings bring additional experimental evidence to recent studies in ARPES ~\cite{Kondo,He}, STS~\cite{Kohsaka},optics~\cite{Bernhard} and transport~\cite{leBoeuf,Florence} all showing that the pseudogap and superconducting phases are in competition.
% even they are intimately conected. 

%Such an s-wave structure for the pseudogap  makes not very probable the performed pairs schem where both the superconducting gap and the pseudogap have the same symmetry.

\section{Nodal versus Antinodal Low Energy Quasiparticle Dynamics}
Here we focus on the very low energy side of the $B_{1g}$ and $B_{2g}$ Raman response functions %and let's see what we can say about the  physical quantities such as the static scattering rate and the spectral weight of quasiparticles 
in and out of the pseudogap phase. 

The inverse of the low energy slope of the electronic Raman response function in the normal state is given by $\zeta(T,\phi)=\frac{1}{(\gamma(\phi))^2 N_{F}(\phi)}. \frac{\Gamma(T,\phi)}{(Z^{*})^2(T,\phi)}$ according to ~\cite{Devereaux2}. $\Gamma$ and  $Z^{*}$ are respectively the static scattering rate and the renormalized spectral weight of the quasiparticles.  $N_{F}$ is defined as the density of states at the Fermi level above T* and $\gamma$ the Raman vertex. %We assume here that $\gamma$ is non temeprature dependent. 
The angle $\phi$ defines a direction in the Brillouin zone with respect to the principal axis. $\phi$ is equal to $0$ and $\pi/4$ for the AN and N directions respectively. 

\begin{figure}[ht!]
\begin{center}
\includegraphics[width=8cm]{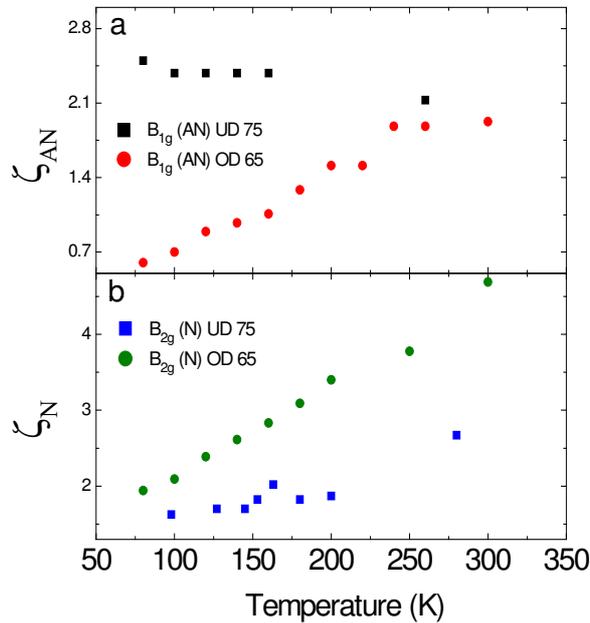}
\end{center}\vspace{-7mm}
\caption{Temperature dependence of (a) $\zeta_{AN}$ and (b) $\zeta_{N}$ for the UD 75 and OD 65 compounds.$\zeta_{AN}$ has been extracted from the Raman spectra in fig.1.} 
\label{fig7}
\end{figure}

Fig. 7-a shows the temperature dependences of $\zeta_{AN}$ for the UD 65 and OD 65 crystals.  $\zeta_{AN}$ exhibits different temperature dependences depending on the doping level. $\zeta_{AN}$ is almost temperature independent for UD 75 while it exhibits a quasi linear temperature dependence for OD 65. %Above $T^*\approx 230~K$,  $\zeta_{AN}$ of UD 75 and OD 65 merge together which signals the end of the pseudogap.  

%$\zeta$ exhibits drastic change according to the nodal and anitnodal direction  in and out of the pseudogap pahse.
%This suggests that quasiparticles are strongly affected by the peudogap in the antinodal region since they are not anymore sensitive to temperature. 
%Asuming Z* non temperature dependent, we conclude that the pseudogap affect drastically the scattering rate which saturates and becomes insensitive to the temperature.We cannot speak anymore of quasiparticle in the antinodal region.The pseudogap altered and destroy quasiparticles at the antinodes.
%However we can notice some differences between the
%In the case of (p=0.22), $\zeta(\pi/4)$ exhibits a linear temperature dependence much more pronounced thant for p=0.22.  
%Even if the slope of $\zeta(\pi/4)$ versus temperature is weaker than for the overdoped but the temperature dependence is the  doping range (p=0.11 to 0.22).
%The low energy nodal quasiparticle dynamic induced by temperature is still there even at low doping level in the pseudogap phase.
%at 300K  in B1g and B2g geometries on $Bi-2212$ single crystals with various doping levels advocate in favor of this scenario. 
%This is effectvely what we experimentally detected from our experiments (see fig.6).  
 
\par
In contrast, $\zeta_N$ for both the OD 65 and UD 75 increases continuously upon heating (see Fig. 7-b). We can also note that the temperature dependence of $\zeta_N$ is stronger for the OD 65 than for the UD 75 compound. Even above $T^*$ ($230~K$),  it subsists  a factor of about $1.6$ between the  $\zeta_N$ values for the two doping levels (see fig.7-b). What does it means ? Does the nodal static scattering rate is larger for OD 65 than for UD 75 or/and the nodal quasiparticle spectral weight is smaller in the OD case than in the UD one ?  

A decrease of $Z^{*}_N$ with doping could be considered as paradoxical and unexpected since an increase of the doping level should a priori enhance the quasiparticle spectral weight. In fact, the  quasiparticles dynamic at the nodes and antinodes have to be considered separately. This dichotomy is clearly seen at $300~K$ (above $T^*$) where the electronic continuum intensities in $B_{1g}$ and $B_{2g}$ exhibit opposite doping depedences \cite{Blanc1}: the intensity of the $B_{2g}$ low energy continuum decreases while the $B_{1g}$ one increases with doping(see fig.8-a and 8-b).

\begin{figure}[ht!]
\begin{center}
\includegraphics[width=14cm]{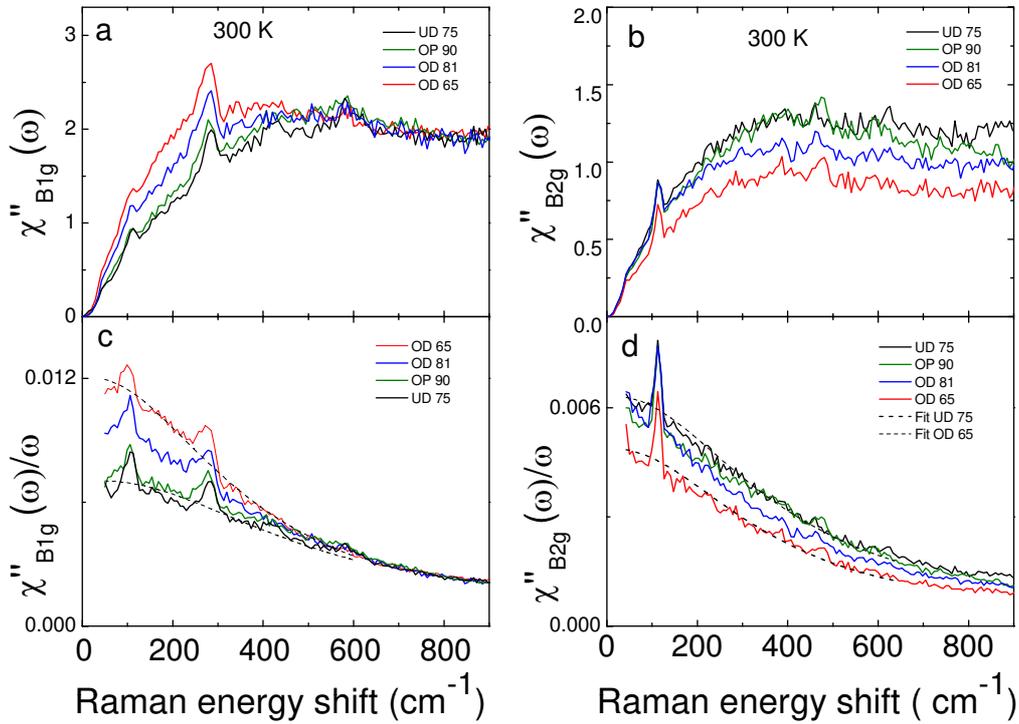}
\end{center}\vspace{-7mm}
\caption{ (a) and (b) $B_{1g}$ and $B_{2g}$ Raman spectra measured at 300 K  for various doping levels; (c) and (d) Raman spectra (a,b) divided by $\omega$. The dash lines correspond to the Drude fits for selected spectra.} 
\label{fig8}
\end{figure} 

\par
In order to emphasize this, we have reported in figs. 8-a and -b the Raman responses  $\chi''(\omega)$ divided by the frequency, $\omega$. According to the Shrastry-Shraiman relation \cite{Shastry} it is related to the real part of the optical conductivity $\chi''(\omega)/\omega$=Re$\sigma(\omega)$. We then observe (see Fig. 8-c and d) a ''Drude'' peak which grows up in the $B_{1g}$ Raman like conductivity spectra whereas it decreases in intensity as the doping level is increased in the $B_{2g}$ geometry.

This behaviour is in contradiction with recent cluster dynamical mean field calculations where the Drude peak of the calculated $B_{2g}$ Raman spectra is seen to grow up and sharpen as the doping level is increased \cite{Lin}. We can notice that the ab plane optical conductivity measurements also exhibit an enhancement of the a Drude peak with doping but optical conductivity is not a k-selective probe \cite{Hwang}. We do not yet understand this discrepancy but we suspect that an ingredient is missing such as a possible reconstruction of the Fermi surface with doping \cite{leBoeuf,Norman2}.  
  %In order to quantify these observations 

We have fitted the $\chi''(\omega)/\omega$ curves with a standard Drude expression assumed to hold at sufficiently low energy \cite{Devereaux2}: $\Gamma(0)(Z^{*})^2/(\Gamma^2(0)+\omega^2)$. $\Gamma(0)$ is the static scattering rate.  This hypothesis makes sense since the scattering rate measured from optical measurements on the same compounds %deduced from the real and imaginary parts of the Raman like conductivity spectra 
is rougthly constant up to $550~cm^{-1}$ \cite{Hwang}. %It has been  obtained from the real part of $\sigma(\omega)=\chi''(\omega)/\omega$ and its 
%Imaginary part has been obtained from Kramers-Kroning transformation \cite{Lobo}. 
The Drude expression (just above) is particulary useful here because it allows us to disentangle between  $\Gamma$ and $Z^{*}$.

%What we cannot  achieve from the low energy slope of the Raman reponse function which is proportionela to the ratio of these both quantities. 

\begin{figure}[ht!]
\begin{center}
\includegraphics[width=8cm]{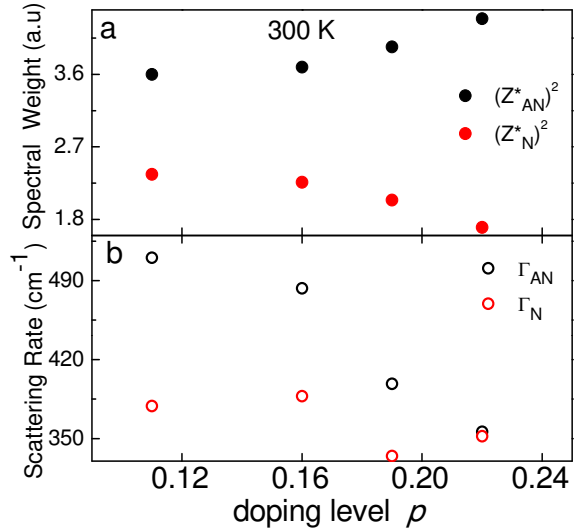}
\end{center}\vspace{-7mm}
\caption{Doping dependence of (a) the quasiparticle spectral weights and (b) the scattering rates in the nodal and antinodal regions. Data have been extracted from the Drude fits of the Raman like conductivity spectra of fig.8.}
\label{fig9}
\end{figure}

The doping dependence of $(Z^{*})^2$ and $\Gamma$ extracted from the Drude fits at $300~ K$ in $B_{1g}$ and $B_{2g}$ geometries are reported in Fig. 9-a and b.  In fig. 9-a, we observe a decrease of $(Z^{*}_N)^2$ of roughtly 40\% with doping while $(Z^{*}_{AN})^2$ increases.  This unexpected decrease of  $(Z^{*}_N)^2$ with doping is also detected from the Drude fits performed on the $B_{2g}$ Raman like conductivity spectra measured at different temperatures %($100~K$ and $200~K$) 
for the UD 75 and OD 65 compounds (see fig. 10-a). 
%( in the doping range  between 0.11 and 0.22) 

%We believe this is a crucial experimental finding not yet enough considered by recent theoretical investigations. 

In Fig. 9-b, $\Gamma_{AN}$ exhibits a clear decrease of about 45\% upon doping: the antinodal quasiparticles are better defined at high doping level than at low one. On the contrary, the nodal scattering rate $\Gamma_{N}$,  is barely doping dependent and shows only  a small decrease of 7\%. Such variations of the static scattering rate are consistent with previous Raman investigations~\cite{Opel,Venturini}. 

%However our scattering rate values are smaller by roughtly a factor 2 with respect to ref.~\cite{Venturini}. the reason is that we have taken into account in our calculation the doping dependence  of $(Z^{*}$. 
\par
From this analysis, it clearly appears that the main contribution responsible for the large difference between $\zeta_N$ between OD 65 and UD 75 (at $300~K$) comes from a decrease of $(Z^{*}_N)^2$  rather than a change in $\Gamma_N$ with doping. 
Notice here that the doping evolutions of the Raman vertex and the density of states $N_{F}$ are unable to explain the increase of $\zeta_N$ with underdoping.  The first one (which is proportional in a first approximation to $t'$, the second nearest-neighbor intraplane hopping integral, is expected to decrease with underdoping ~\cite{Kordyuk} while $N_{F}$ around the nodes is expected to be almost constant in this doping range ($0.11-0.22$) ~\cite{Kordyuk}. 

%The decrease of $Z^{*}_N$ as the doping level increases between $p=0.11$ and $p=0.22$ appears to be experimentally real and robust. 
 \par
We would like to mention here that the  $Z^{*}_N$ increase with underdoping is perfectly consistent with our previous investigations in the superconducting state. We have shown that the nodal slope of the $B_{2g}$ superconducting Raman response $\frac{N_F (Z^{*}_{N})^2}{v_{\Delta}}$ is almost constant ~\cite{Blanc1,LeTacon}. The most likely scenario for this is that $v_{\Delta}$, the nodal slope of the superconducting gap increases with underdoping~\cite{Sacuto,Blanc2}. This implies that $Z^{*}_N$ increases with underdoping. 

%  slope with underdoping This  adovcates in favour of an  increase of the nodal slope of the superconducting gap with underdoping since the ratio has been shown to be almost constant  in our previous works ( Le tacon et S.Blanc et al PRB 1).

\begin{figure}[ht!]
\begin{center}
\includegraphics[width=8cm]{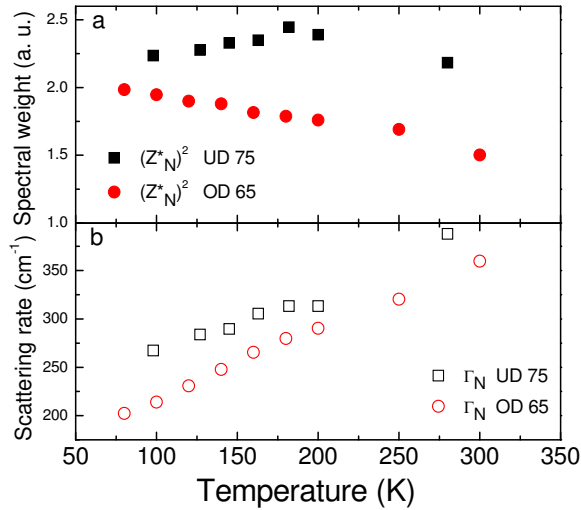}
\end{center}\vspace{-7mm}
\caption{Temperature dependence of (a) the quasiparticles spectral weight and (b) the scattering rates for the UD 75 and OD 65 crystals.} 
\label{fig10}
\end{figure}

Such a strange doping evolution of $Z^{*}_N$ is not yet elucidated. However we believe that Fermi surface disturbances could induce this kind of behaviour. 
% active in both the normal and superconducting state. 

\par
By applying the Drude fit procedure on the Raman like conductivity spectra, we can also extract the temperature dependence of both  $(Z^{*}_N)^2$ and $\Gamma_N$. This is shown in fig. 10. Surprisingly we find that $Z^{*}_N$ exhibits a weak temperature dependence for UD 75 (7\%) and a larger one of about 30\% for OD 65. We do not have at present an explanation for this behaviour. We also observe as already mentioned that the $(Z^{*}_N)^2$ values (in the temperature range considered) are larger for the UD 75 than for the OD 65 compounds. In fig. 10-b, the static scattering rate, $\Gamma_N$, shows a quasi linear temperature dependence for  both the UD 75 and OD 65 compounds. It is quite similar to the temperature dependence of the scattering rate obtained from earlier Raman experiments ~\cite{Opel}. The temperature dependence of $\Gamma_N$ is quite robust in the doping range $p=0.11$ and $0.22$. We have to investigate Raman measurements at higher doping level to detect the $T^2$ behaviour seen from quantum oscillations measurements on strongly overdoped compounds (OD 15K) \cite{Hussey}. 

%We have to perform Raman measurements on more overdoped compounds.  

\section{Conclusion}

The pseudogap manifests itself in different manners in the $B_{1g}$ and  $B_{2g}$ Raman response functions of underdoped Bi-2212 single crystals. 
In the $B_{1g}$ spectra, it corresponds to a strong depletion of the low energy electronic continuum as the temperature decreases. This has permitted us to identify both the pseudogap temperature and the pseudogap energy as a function of doping level. The effect of the pseudogap in the $B_{2g}$ Raman spectra  is more subtle. The depletion of the electronic continuum settles at a higer energy range than in the $B_{1g}$ geometry. Moreover, superconductivity brings supplementary electronic states inside the energy range where the depletion develops. This is in contradiction with a preformed pairs scenario and lead us to suggest different structures for the pseudogap and the superconducting gap~\cite{Sakai}: a d-wave superconducting gap and an s-wave anisotropic pseudogap. We show that the antinodal and nodal quasiparticle dynamics are distinct. The nodal scattering rate exhibits a quasi linear temperature dependence both inside and outside the pseudogap phase while the antinodal scattering rate is strongly altered and becomes non temperature dependent inside the pseudogap phase. This reveals the robutness of the nodal quasiparticles with doping range $p=0.1-0.2$. Finally, we find that the nodal quasiparticles spectral weight increases with underdoping while the antinodal one decreases. We believe these observations deserve to be more deeply investigated.

\subsection{Acknowledgments}

We thanks S. Sakai, M. Civelli, A. Georges, A. Millis, G. Orso and R. Lobo for very fruitful discussions.

\section{References}
%%%%%%%%%%%%%%%%%%%%%%%%%%%%%%%%%%%%%%%%%%%


\begin{thebibliography}{9}
\bibitem{note} $Z^{*}$ is equivalent to $Z\Lambda$ defined in ref.~\cite{LeTacon}.
\end{thebibliography}
\end{document}